\documentclass[aps,pre,twocolumn,showpacs]{revtex4-1}
\usepackage{graphicx}
\usepackage{bbold}
\usepackage{epsfig}
\usepackage{subfigure}
\usepackage{amsmath}
\newcommand{\eqa}{\begin{eqnarray}}
\newcommand{\eqe}{\end{eqnarray}}
\newcommand{\bq}{{\bf q}}
\newcommand{\br}{{\bf r}}
\newcommand{\bQ}{{\bf Q}}
\newcommand{\s}{{\sigma}}
\newcommand{\bs}{{\bar{\sigma}}}

\begin{document}

\title{Interaction-driven capacitance in graphene electron-hole double layer in strong magnetic fields}

\author{Bahman Roostaei}

\affiliation {Department of Physics, Indiana University-Purdue University Indianapolis, Indiana 46202, USA}
\affiliation {School of Physics, Institute for Research in Fundamental Sciences (IPM), P.O. Box 19395-5531, Tehran, Iran}

\date{\today}
\begin{abstract}
Fabrication of devices made by isolated graphene layers has opened up possibility of examining highly correlated states of electron systems in parts of their phase diagram that is impossible to access in their counterpart devices such as semiconductor heterostructures.  An example of such states are graphene double mono-layer electron-hole systems under strong magnetic fields where the separation between layers can be adjusted to be as small as one magnetic length with interlayer tunneling still suppressed. In those separations it is known that correlations between electrons and holes are of crucial importance and must be included in determination of observable quantities. Here we report the results of our full numerical Hartree-Fock study of coherent and crystalline ground states of the interacting balanced electron-hole graphene systems in small and intermediate separations with each layer occupying up to four lowest lying Landau levels. We show that in the Hartree-Fock approximation the electrons and holes pair to form a homogeneous Bose-condensed (excitonic) state while crystalline states of such exciton systems remain incoherent at intermediate layer separations. Our results of calculation of capacitance of such states as a function of inter-layer separation and filling factor provides quantitative and qualitative signatures that can be examined in real experiments. We show that the capacitance of some crystallized states as well as uniform coherent states are significantly enhanced compared to geometrical values solely due to Coulomb interactions and quantum corrections.
\end{abstract}
\maketitle

\section{Introduction}
The study of electron gas in graphene systems has recently taken a tremendous advantage from capacitance measurements\cite{Skinner2013,Skinner2013a,Yu2013}. This is because transport measurements are not able to reveal certain states of matter due to disorder effects. In fact quantum corrections to capacitance of single layer graphene under the influence of quantizing magnetic fields has been predicted and experimentally shown\cite{Skinner2013,Skinner2013a,Yu2013} to result in large enhancement compared to geometrical value among other features which are not yet fully understood.

In a typical experiment to measure capacitance of an electron gas, the system is placed near a parallel metal electrode gate\cite{Yu2013}. This system has been discussed previously by several authors\cite{Skinner2013,Skinner2013a,skinner2010} using the idea of image charges: The two dimensional electron gas plus the electrode gate can be thought of as a gas of dipoles made out of electrons with their image charges\cite{skinner2010}. According to those authors dipoles in this gas interact more weakly and as the result layers allow easier accommodation of charges which means an enhancement of capacitance. Study of phases and the capacitance of dipoles in an electron-hole bilayer system is therefore the focus of this article.


Generally interaction between electrons in one layer and holes in the other layer could result in non-trivial quantum states with broken symmetries\cite{E-H}. Historically the problem of electron-hole bilayer system has not been theoretically investigated as much as the problem of electron-electron system. This has been partly because of the difficulty of preparing such systems in semiconducting heterostructures with identical electron and hole properties. With graphene available we have now an ideal material in which the polarity of carriers are adjustable \textit{in situ} and there is virtually perfect symmetry between electron and hole bands at typical carrier densities. This fact makes graphene an ideal candidate to study various two dimensional quantum states formed by interacting electron and hole layers. Experimental studies of those states can be done for example by observation of the capacitance of the electrode-graphene sheet device\cite{Skinner2013,Skinner2013a,Yu2013} or by forming double electron-hole graphene layers either using electrostatic gates\cite{Yu2013} or by optical excitations\cite{Butov2004}.

In studies of the electron-electron systems electron filling factor of the two layers are tuned to be the same or different. The balanced electron-hole system in which the electron filling factor is the same as the hole filling factor is equivalent to the imbalanced electro-electron system if one performs a particle-hole transformation in the hole layer. It is now established\cite{Fertig89,stablished-state} that when the total electron filling factor in such system is equal to one the true ground state is a uniform integer quantum Hall state. In such case the effect of imbalance between the electron filling factor of the two layers has been studied\cite{imbalance}. It has been found out that the imbalance moves the phase boundary between the coherent state and incoherent state by enhancing the coherent state of the double layer system. The coherence here means each electron is in symmetric state between the two layers even at the vanishing limit of tunneling. Early works\cite{Chen1991} on balanced electron-hole systems indicated the appearance of roton minimum in the collective excitation spectrum of the system as the layer separation is lowered. This suggested the existence of an instability of the state of the system toward the formation of a charge-density-wave state or a Wigner crystal state. We will compare the ground states that we find in this article with those findings. Also studies of the balanced electron-electron system at total filling factor less than one were indicating the existence of Wigner crystal ground states\cite{Zheng-Fertig} however unfortunately full study of a crystal state of imbalanced electron-electron system was never performed to the best of the author's knowledge and is still missing. That is another main motivation of this article. The problem of Wigner crystal states in balanced electron-electron system\cite{Cote1992,Narasimhan1995,Csaba2014} and balanced electron-hole system exactly match when the filling factor of electron and hole layers are both at 1/2 and interlayer tunneling does not exist. One must note however that even if there is no tunneling between the layers in the electron-hole system, there is still quantum coherence expected to happen when excitons (we use dipoles and excitons interchangeably throughout this article) develop a non-zero average phase throughout the system. This phase will be defined when we introduce the density matrix of the system in section \ref{section-HF}.

Historically in a single layer gas of charged carriers at zero magnetic field and at low enough densities it is well known\cite{Wigner} that the carriers form a crystalline state known as Wigner crystal assuming they are not massless. This is because as a function of density $n$ in two dimensions kinetic energy of carriers behaves as $n$ while Coulomb energy behaves as $n^{1/2}$. This results in the Coulomb energy being dominant at low enough densities and the carriers find a low energy crystalline state. For low densities of a dipole gas on the other hand where the Coulomb interaction scales as $r^{-3}$ ($r$ being the distance between dipoles) this argument breaks down and the crystallization seems impossible. However application of a strong perpendicular magnetic field freezes electron's and hole's (therefore the dipole's) kinetic energy and localizes their wavefunction. In this case, quantum fluctuations are reduced and formation of a crystal is indeed possible. This crystal is characterized by two length scales, the magnetic length $\ell_B=\sqrt{\hbar/eB}$ and dipole arm length $d$ or the electron-hole layer separation. Magnetic length represents the average spatial separation between carriers, this time even more controllable thanks to its dependance on the magnetic field.

In this article we systematically investigate all possible crystalline ground states of an electron-hole system and calculate their associated capacitance. In a regime where electrons or holes in separate isolated mono-layers of graphene under strong magnetic fields form Wigner crystal ground states we can ask the question of what phases form in the ground state when the two such crystals (one electron and the other holes) are brought adjacent to each other? In today's state of the art technology it is indeed possible to achieve such arrangement by fabricating stacks of graphene layers with separations as small as one magnetic length\cite{Geim2013}. In the presence of magnetic field, since only the ratio $d/\ell_B\propto \sqrt{B}$ is physically important tuning the magnetic field covers all desired inter-layer separations. On the other hand in small separations interlayer tunneling is still highly suppressed contrary to semiconductor counterparts, thanks to the highly resistive insulating barriers (such as hexagonal boron nitride).

At separations comparable to or smaller than a magnetic length ($d \sim \ell_B$) it is expected that electrons and holes in the two separate layers form exciton pairs in ground state. This exciton gas is then predicted to break U(1) symmetry (associated with the phase of the pair wave function) and the system undergo a phase transition into an excitonic condensate state\cite{Yoshioka1990} in which excitons as bosons condense into the lowest lying bosonic state. Therefore we need to consider crystallization as well as U(1) symmetry breaking in ground state investigations. Such analysis has been done in the past\cite{Chen1991} only for unidirectional states and only at lowest Landau level.

In this article we assume the electrons and holes with the same filling factor ($\nu_e=\nu_h=\nu$) are confined in the lowest $N=0,1,2,3$ Landau levels with the last level partially filled. We specifically consider dipolar states in which electrons and holes gain the maximum attraction energy by having the same local density. We show that we recover the uniform excitonic state at small interlayer separations, $d < \ell_B $ found previously\cite{Yoshioka1990}. Throughout our calculations we ignore the inter-Landau level transitions and layer thickness. We then use the Hartree-Fock (HF) approximation to obtain the ground state density matrix and associated energies as a function of layer separation and filling factor. The central findings of our investigations are as follows:


\textit{Ground State Phase Diagram} - In the phase space of filling factor and layer separation the crystalline ground states is formed at $d \gtrsim \ell_B$ and $\nu_T=\nu_e+\nu_h \lesssim 0.8$ and the crystal type becomes more and more anisotropic in higher Landau levels. Also at higher Landau levels the crystalline state phase boundary moves further toward smaller separations. We also do not find states in which crystals of excitons develop a uniform or modulated phase (broken U(1) symmetry) in ground state in our approximations. We will discuss this issue in the later sections of this work.

\textit{Capacitance} - Our calculations of capacitance associated with the crystalline states on the other hand indicates a remarkable enhancement compared to geometrical value specially for lower Landau levels while at higher levels ($N=1,2,3$) the corrections for some filling factors reduce the capacitance. In all transitions from crystalline state to uniform excitonic state, there is a jump (indicating a sharp change in reality) in capacitance value.

This paper is organized as follows: In section \ref{section-HF} we explain the basic formulation of HF approximation that has been used to calculate the density matrix and energy of the system at zero temperature. This section is brief and only explains the main lines of argument in deriving the HF equations because this method has been frequently used in the past in the literature for single and double layer electron systems in two dimensions under strong magnetic field. In section \ref{section-num} we explain our numerical results for both phase diagram and capacitance. We also explain in this section, the quantum corrections to capacitance obtained by HF approximation. Finally in section \ref{section-last} we discuss our findings for both the phase diagram and capacitance and briefly explain possible extensions of the problem and open questions.

\section{Hartree-Fock Approximation}\label{section-HF}
\subsection{Density and Energy}
In this section we briefly explain the Hartree-Fock approximation that we use to calculate the energies and density matrices associated with different crystalline states. We start by introducing the Hamiltonian for electrons and holes in adjacent sheets of graphene around the Dirac points under strong magnetic field. The low-energy Hamiltonian for free electrons or holes in ${\bf K}$ valley of the graphene sheet is given by\cite{Zhang2007}:
\eqa\label{G-Ham}
H_{\bf K}=v(p_x\tau_x+p_y\tau_y)
\eqe
where $\tau_x$ and $\tau_y$ are Pauli matrices in the space of sublattices A, B within a unit cell\cite{CastroNeto2009}. The parameter $v$ is the Fermi velocity of electrons. We assume the valley splitting to be negligible and since the inter-valley scattering is small (of the order of $a/\ell_B$ where $a$ is the graphene lattice constant) in strong magnetic fields we consider electrons and holes are simply in one valley and ignore the valley degree of freedom. Also we assume Zeeman splitting of Landau levels to be large enough that the spins are completely polarized and the spin degree of freedom is frozen.

In the presence of a magnetic field the free-particle Hamiltonian includes a gauge potential  ${\bf p}\rightarrow {\bf p} \pm e {\bf A}$ where plus(minus) sign is for electrons (holes) and we assume $e>0$. In the case that we are considering throughout this article we assume the magnetic field is uniform across the sample ${\bf B}=\nabla\times {\bf A}=B\hat{z}$. In the Landau gauge we have ${\bf A}=B x\hat{y}$ which results in the Hamiltonian (\ref{G-Ham}) to take the form:
\eqa
H_{\bf K}={\sqrt{2}\hbar v\over \ell_B}\left(
                                          \begin{array}{cc}
                                            0 & c_{k} \\
                                            c^\dagger_{k} & 0 \\
                                          \end{array}
                                        \right)
\eqe
where $c_k=-i[\ell_B\partial_x+(x/\ell_B-k\ell_B)]/\sqrt{2}$ is the lowering operator for electrons and holes, $k\ell_B^2$ is the guiding center position which counts the degeneracy of each Landau level and $\left[c_k,c^\dagger_{k'}\right]=\delta_{k,k'}$. The eigenvalues of this Hamiltonian are given by $E_N=\pm\hbar v \sqrt{2|N|}/\ell_B$ in which $N$ is the Landau level index and the eigenfunctions are:
\eqa
\langle {\bf r}|{\bf K}, Nk\rangle={1\over \sqrt{2L_y}}e^{iky}\left[
                                                                \begin{array}{c}
                                                                  \text{sgn}(N)\phi_{|N|-1}(x-k\ell_B^2) \\
                                                                  \phi_{|N|}(x-k\ell_B^2) \\
                                                                \end{array}
                                                              \right]
\eqe
For $N\neq 0$. For $N=0$:
\eqa
\langle {\bf r}|{\bf K}, 0k\rangle={1\over \sqrt{2L_y}}e^{iky}\left[
                                                                \begin{array}{c}
                                                                  0 \\
                                                                  \phi_{0}(x-k\ell_B^2) \\
                                                                \end{array}
                                                              \right].
\eqe
In the above $\phi_N(x)$ are the simple harmonic oscillator eigenfunctions and $L_y$ is the sample length in the $y$ direction. From now on we will use units such that $\hbar=1$. The two dimensional plane of the graphene sheet is described by $\{x,y\}$ coordinates and the center of top and bottom sheets are located at $z=+d/2$ and $z=-d/2$. In the limit of strong enough magnetic field it is experimentally possible to concentrate the electron/hole system in single or several lowest Landau levels. The Hamiltonian for the interacting electron-hole system consists of kinetic and Coulomb energies. We use index $\sigma=e(h)$ for various functions to denote electrons(holes). In the single particle basis after Fourier transformation the full Hamiltonian including the Coulomb interactions is:
\eqa
\hat{H}&=& N_\phi\sum_{N,\sigma}E_N \hat{\rho}^{\sigma\sigma}_{NN}(0)+\nonumber\\
&+& {N_\phi\over 4\pi\ell_B^2}\sum_{\{N\}}\sum_{\{\sigma\},\bq}V_{\sigma_1\sigma_2}(\bq){\mathcal F}_{N_1N_4}(\bq){\mathcal F}_{N_2N_3}(\bq)\times\nonumber\\
&\times& \hat{\rho}^{\sigma_1\sigma_1}_{N_1N_4}(-\bq)\hat{\rho}^{\sigma_2\sigma_2}_{N_2N_3}(\bq)
\eqe
where $N_\phi=A/(2\pi\ell_B^2)$ is the number of flux quanta in the area $A$ of the sheet. The functions $V_{ee}(\bq)=V_{hh}(\bq)=2\pi e^2/(\epsilon q)$ are the Fourier transform of intra-layer Coulomb repulsion and $V_{eh}(\bq)=-V_{ee}(\bq)\exp(-qd)$ is the inter-layer Coulomb attraction between electrons and holes. The Fourier transform of density matrix elements are defined as:
\eqa
\hat{\rho}^{\sigma\sigma}_{NN'}(\bq)={1\over N_\phi}\sum_{k,k'}e^{-(i/2)q_x(k+k')\ell_B^2}c_{\sigma N k}^\dagger c_{\sigma N' k'}\delta_{k,k'+qy}\nonumber\\
\eqe
and for electron-hole pairing operator:
\eqa
\hat{\rho}^{eh}_{NN'}(\bq)={1\over N_\phi}\sum_{k,k'}e^{-(i/2)q_x(k+k')\ell_B^2}c_{e N k}^\dagger c_{h N' -k'}^\dagger\delta_{k,-k'+qy}\nonumber\\
\eqe
in which $c_{\sigma N k}^\dagger$($c_{\sigma N k}$) creates (annihilates) an electron (hole) in the Landau level $N$ with the state $|N,k\rangle$. Recall that single-particle wave function for holes are obtained by complex-conjugating that of electrons. Finally the form factors are defined as:
\eqa
{\mathcal F}_{NN'}(\bq)&=& \delta_{N,0}\delta_{N',0}F_{N,N'}(\bq)+\nonumber\\
&+&{1\over \sqrt{2}}\delta_{NN',0}\delta_{N+N'\neq 0}F_{N,N'}(\bq)+\nonumber\\
&+& {1\over 2}\theta(|N|)\theta(|N'|)[F_{|N|,|N'|}(\bq)+\nonumber\\
&+&\text{sgn}(NN')F_{|N|-1,|N'|-1}(\bq)]\nonumber\\
\eqe
where $\theta(x)$ is the Heaviside function. This form factor is a linear combination of contributions from the wave functions of the two inequivalent lattice sites,

\eqa
F_{N\geq N'}(\bq) &=& \left[{N'!\over N!}\right]^{1/2}\left[{(-q_y+iq_x)\ell_B\over \sqrt{2}}\right]^{N-N'}\times\nonumber\\
&\times& \exp\left[{-q^2\ell_B^2\over 4}\right]L_{N'}^{N-N'}\left[{q^2\ell_B^2\over 2}\right]
\eqe
for $N'\leq N$ and $L_N^a(x)$ is the generalized Laguerre polynomial. Note that from the above we see that $F_{NN'}(\bq)=\left[F_{N'N}(-\bq)\right]^*$.

 The derivation of the Hamiltonian in HF approximation in terms of above density matrices has been discussed in great detail in the past literature\cite{method}. Throughout this article we ignore inter-Landau level transitions. This is justified because the ratio of the Landau level gap $E_{N+1}-E_N\approx\sqrt{2}v/\epsilon \ell_B $ to inter-particle Coulomb interaction energy $e^2/\epsilon\ell_B$ is larger than one ($\approx 3.2$ for values of $\epsilon\approx 2-5$ taken from literature\cite{Dahal2006}). This along with the HF approximation will simplify the Hamiltonian to the following:
\eqa\label{HF-Ham}
H_{HF}&=& {N_\phi e^2\over \epsilon\ell_B}\sum_{\sigma,\bQ}W_N^\sigma(\bQ)\hat{\rho}_N^{\s\s}(\bQ)+\nonumber\\
&-& {N_\phi e^2\over \epsilon\ell_B}\sum_{\sigma,\bQ}\left[H_N^{\bs\s}(\bQ)\hat{\rho}_N^{\s\s}(\bQ)+X_N^{\s\bs}(\bQ)\hat{\rho}_N^{\bs\s}(\bQ)\right]\nonumber\\
\eqe
where we have already assumed the density matrix is nonzero only at certain wavevectors belonging to a group of reciprocal lattice vectors (RLV) $\{\bQ\}$ associated with a crystal of choice. The Hartree-Fock potentials in the above are obtained as:
\eqa
W_N^\sigma(\bQ)&=& \left[{E_N\over e^2/\epsilon \ell_B}\delta_{\bQ,0}+H_N^{\s\s}(\bQ)-X_N^{\s\s}(\bQ)\right]\\
H_N^{\s\s'}(\bQ) &=&{1\over Q\ell_B}\exp\left[-Q^2\ell_B^2/2-Qd_{\s\s'}\right]\times\nonumber\\
&\times& |{\mathcal F}_{N,N}(\bQ)|^2\rho_N^{\s\s'}(-\bQ)\\
X_N^{\s\s'}(\bQ) &=& \int_0^{\infty} dx \exp\left[-x^2/2-x {d_{\s\s'}\over\ell_B}\right]\times\nonumber\\
&\times& |{\mathcal F}_{N,N}(x)|^2J_0(Q\ell_B x)\rho_N^{\s\s'}(\bQ)
\eqe
in which $d_{\s\s}=0$, $d_{\s\bs}=d$, $\bs=-\s$ and $J_0(x)$ is the Bessel function of the first kind. Because of the existence of a uniform neutralizing background charge density close to both layers there is an extra uniform capacitive term: $H_N^{\s\s}(0)-H_N^{\bs\bs}(0)=d\nu\ell_B$ and $H_N^{\s\s}(0)+H_N^{\bs\bs}(0)=0$. Also in the above $\rho_N^{\s\s'}(\bQ)=\langle \hat{\rho}_{NN}^{\s\s'}(\bQ)\rangle$.
 In order to find the density matrix self consistently one needs to introduce the $2\times 2$ Green's function matrix defined as follows:
\eqa
G_N(k_1,k_2;\tau)=-\langle T a_{Nk_1}(\tau)a^{\dagger}_{Nk_2}(0)\rangle
\eqe
in which the vector $a^\dagger_{Nk}=\left(c^\dagger_{Ne,k},c_{Nh,-k}\right)$. The Fourier transform of such function is also obtained by:
\eqa
G_N(\bQ,i\omega_n) &=& {1\over N_\phi}\sum_{k_1,k_2}\int_0^{\beta} d\tau e^{-iQ_x(k_1+k_2)\ell_B^2/2+i\omega_n\tau}\nonumber\\
&\times& \delta_{k_2,k_1-Q_y}G_N(k_1,k_2;\tau)
\eqe
where $\omega_n$ is a Matsubara frequency and $\beta = 1/k_B T$ in the inverse temperature. Throughout this article we use the limit of $T\rightarrow 0$ for ground state. Using the HF Hamiltonian the equations of motion for Green's functions can be obtained. The electron and hole density matrices are then determined from equal-time limit $\tau\rightarrow 0^-$ of the Green's function matrix. The equation of motion is as follows:
\eqa\label{eq-motion}
 \delta_{\bQ,0}I&=&\left[
  \begin{array}{cc}
    i\omega+\mu & 0 \\
    0 & i\omega-\mu \\
  \end{array}
\right]G_N(\bQ,i\omega)+\nonumber\\
&-& \sum_{\bQ'}\mathcal{M}(\bQ-\bQ')e^{i\bQ\times\bQ'\ell_B^2/2}G_N(\bQ',i\omega)
\eqe
where the self energy matrix $\mathcal{M}$ is defined as follows:
\eqa
\mathcal{M}(\bQ)=\left[
                                     \begin{array}{cc}
                                       \Sigma_{ee}(\bQ-\bQ') & \Sigma_{eh}(\bQ-\bQ') \\
                                       \Sigma_{he}(\bQ-\bQ') & -\Sigma_{hh}(\bQ-\bQ') \\
                                     \end{array}
                                   \right]
\eqe
with the elements:
\eqa
\Sigma_{ee}(\bQ) &=& \left[H_N^{ee}(\bQ)-X_N^{ee}(\bQ)\right]\rho_N^{ee}(-\bQ)\nonumber\\
&-& H_N^{eh}(\bQ)\rho_N^{hh}(-\bQ)\\
\Sigma_{eh}(\bQ) &=& -X_N^{eh}(\bQ)\rho_N^{eh}(-\bQ)\nonumber\\
\eqe
and the other two elements are obtained simply by $e\leftrightarrow h$. The solution to this equation can be obtained by diagonalizing the self-energy matrix and using the eigenvectors $\lambda^\dagger_j(\bQ)=\left[V^*_j(\bQ),U^*_j(\bQ)\right]$ and associated eigenvalues $\Omega_j$ as follows:
\eqa\label{eigval-eq}
\sum_{\bQ'} &&\left[\mathcal{M}(\bQ-\bQ')-\mu\tau_z\delta_{\bQ,\bQ'}\right] e^{i\bQ\times\bQ'\ell_B^2/2}\lambda_j(\bQ')=\nonumber\\
&=&\Omega_j\lambda_j(\bQ).
\eqe
The solution to the above equation will be:
\eqa\label{green-eq}
G_N(\bQ,i\omega)=\sum_j {\lambda_k(\bQ)\lambda^\dagger_k(0)\over i\omega - \Omega_j}
\eqe
The chemical potential is obtained during the self-consistent calculation by the constraint that:
\eqa
\rho_N^{ee}(0)=\nu_e=\rho_N^{hh}(0)=\nu_h=\nu.
\eqe
From now on we will use $\nu$ instead of $\nu_e$ or $\nu_h$ and define $\nu_T=\nu_e+\nu_h$. After the density matrix solutions are obtained the HF energy of that state can be calculated using the expectation value of Hamiltonian (\ref{HF-Ham}).
Finally the real space profiles of the density matrix is obtained as:
\eqa
\rho_N(\br)={1\over 2\pi\ell_B^2}\sum_\bQ \rho_N(\bQ){\mathcal F}_{NN}(\bQ)e^{i\bQ\cdot\br}
\eqe

Solutions to the above equations are of two general types: crystalline and uniform states. For the crystalline states translational symmetry of the original Hamiltonian is broken. In this situation the density matrix $\rho$ is non-zero only at certain crystal points ${\bf R}$ or at corresponding RLV's $\bf Q$: $\rho(\bQ)\neq 0$.

There is another symmetry associated with the state of the electron-hole system: the energy of each state depends only on absolute value of the density matrix. The off-diagonal part of the density matrix $\rho_{eh}$ can be in principle a complex number. This element indicates the pairing of the electrons and holes. A non-zero off-diagonal density matrix element indicates excitons (electron-hole pairs) have been formed. The energy of such state would be invariant as the phase of the complex number changes. This is called U(1) symmetry. Basically there is no reason the phase is the same throughout the system. However because of the inter-particle interactions the exciton gas may find a lower energy by breaking this symmetry and choose a uniform or modulated phase throughout the system. This is a state with broken U(1) symmetry.

We call the crystalline states with U(1) symmetry broken \textit{Coherent Wigner Crystal} and denote them by WCC irrespective of the type of the crystal. Those states are crystals of excitons in which the excitons are part of a condensate state as well. For such states $\rho_{\s\s}(\bQ\neq 0)\neq 0$ and $\rho_{\s\overline{\s}}(\bQ)\neq 0$.

It is naturally expected to consider \textit{Incoherent Wigner Crystals} (WC) as states in which only the translational symmetry is broken but not U(1) symmetry. In such states $\rho_{\s\s}(\bQ\neq 0)\neq 0$ and $\rho_{\s\overline{\s}}(\bQ)=0$ where interacting electron-hole \textit{dipoles} have formed crystals.

Last but not least, uniform states where translational symmetry is not broken are possible. For those states $\rho(\bQ\neq 0)=0$. In particular there are solutions in which the pairing has indeed happened and U(1) symmetry is broken: $\rho_{eh}(0)\neq 0$. For those solutions excitons have formed a condensate state. More precisely for a \textit{uniform excitonic condensate} state (UE) we have: $\rho_{ee}(\bQ)=\rho_{hh}(\bQ)=\nu\delta_{\bQ,0}$ and $\rho_{eh}(\bQ)=\alpha\delta_{\bQ,0}$ in which $\alpha$ can be determined self-consistently.

Later on we will present our HF numerical results indicating that the WCC states are always slightly higher in energy than WC states or uniform density states.

In section \ref{section-num} we explain in details all the WC and WCC states that we find from solving the HF equation and we compare their energies at different parts of the phase diagram $(\nu_T,d/\ell_B)$ to find the HF ground state.
\subsection{Capacitance}

In order to calculate the capacitance of a system of charged particles one needs to notice that the finiteness of the density of states of the system must be taken into account. In other words we need to be careful about the change of the chemical potential of the system by changing the number of particles. This point is not relevant in classical systems such as two perfect metal electrodes where we approximately consider their density of states is infinite. The capacitance of system of charged particles is defined by the general formula:
\eqa\label{C-define}
C^{-1}={d^2 U\over dQ^2}
\eqe
in which $U$ is the thermodynamic internal energy of the system. When we consider a system of electron gas adjacent to a classical metal electrode the change in the internal energy of the system at constant temperature and pressure is:
\eqa
dU=dU_e+\mu dN
\eqe
in which $U_e$ is the electrostatic energy, $\mu$ is the chemical potential of the system and $N$ is the number of particles. Using the definition (\ref{C-define}) we find the capacitance of the system \textit{per unit area} to be\cite{skinner2010}:
\eqa
c^{-1}={d\over\epsilon}+{1\over e^2}{d\mu\over dn}
\eqe
where $n$ is the particle density. Note that the first term will be correct only at the limit of perfectly screening metal electrode where the electric field between electron gas and the electrode is uniform otherwise one needs to use the original equation (\ref{C-define}) to obtain the total inverse capacitance. In this article we indeed use this original equation. With those considerations the \textit{Quantum  Capacitance} is defined as $e^2dn/d\mu$ and represents all the corrections for deviation from classical (uniform electric field) standard value. Often in literature an effective thickness is introduced $d^*=\epsilon/c$ which can be written as $d^*=d+d_Q$ where $d_Q$ is called the \textit{quantum capacitance length} (QCL).

In the case of a low density electron gas in a quantum well and zero perpendicular magnetic field it is very well known that the thermodynamic density of states ($dn/d\mu$) can be \textit{negative}\cite{Eisenstein1994} because of strong positional correlations between electrons. This means a negative quantum capacitance length or an enhanced capacitance compared to geometrical value.

We can use our results for the HF energy of ground state of the electron-hole system to calculate the quantum capacitance length at $N=0,1,2,3$. Using the calculated HF energies the quantum capacitance length can be calculated as follows:
\eqa\label{QCL-energy}
d^*={\ell_B\over 2}{d^2\over d\nu^2}\left[{\nu E(\nu)\over e^2/\epsilon\ell_B}\right].
\eqe
where $E(\nu)$ is the energy per electron-hole pair.

 For UE states it is possible to find $d_Q$ analytically since we know $E(\nu)$ analytically\cite{Yoshioka1990}:
\eqa\label{E-uniform}
E(\nu;N)=d{\nu}-{\nu}V_{ex}(N)-\left(1-{\nu}\right)V^d_{ex}(N)
\eqe
in which the intra-layer and interlayer exchange energy for filling factor $\nu$ in $N$-th Landau level are respectively:
\eqa\label{Vx}
V_{ex}(N)=\int_0^{\infty} |B_N(x)|^2 e^{-x^2/2} dx
\eqe
and:
\eqa\label{Vxd}
V_{ex}^d(N)=\int_0^{\infty} |B_N(x)|^2 e^{-x^2/2-x d} dx
\eqe
where:
\eqa\label{B}
B_N(x)={1\over 2}\left[L_N^0\left({x^2\over 2}\right)+L_{N-1}^0\left({x^2\over 2}\right)\right]
\eqe
for $N\neq 0$ and $B_0(x)=1$. The first term in Eq. (\ref{E-uniform}) is the uniform contribution from direct Coulomb interaction.
In the second part of the next section where we present our numerical results for capacitance where we will see how the enhanced capacitance of the electron-hole system changes its behavior as a function of interlayer separation, filling factor and Landau level index.
\section{Results}\label{section-num}
In this section we present the results of our numerical calculations based on the approximations explained in the previous section. We first present examples of the behavior of HF energy of various ground state crystal structures and then present the overall phase diagram. After that we present the quantum corrections to capacitance of the electron-hole double layer system calculated using the energies and densities presented in the first part.

\begin{figure}[t]
     \begin{center}
        \subfigure[]{%
            \label{Genc005-n0}
            \includegraphics[scale=.4]{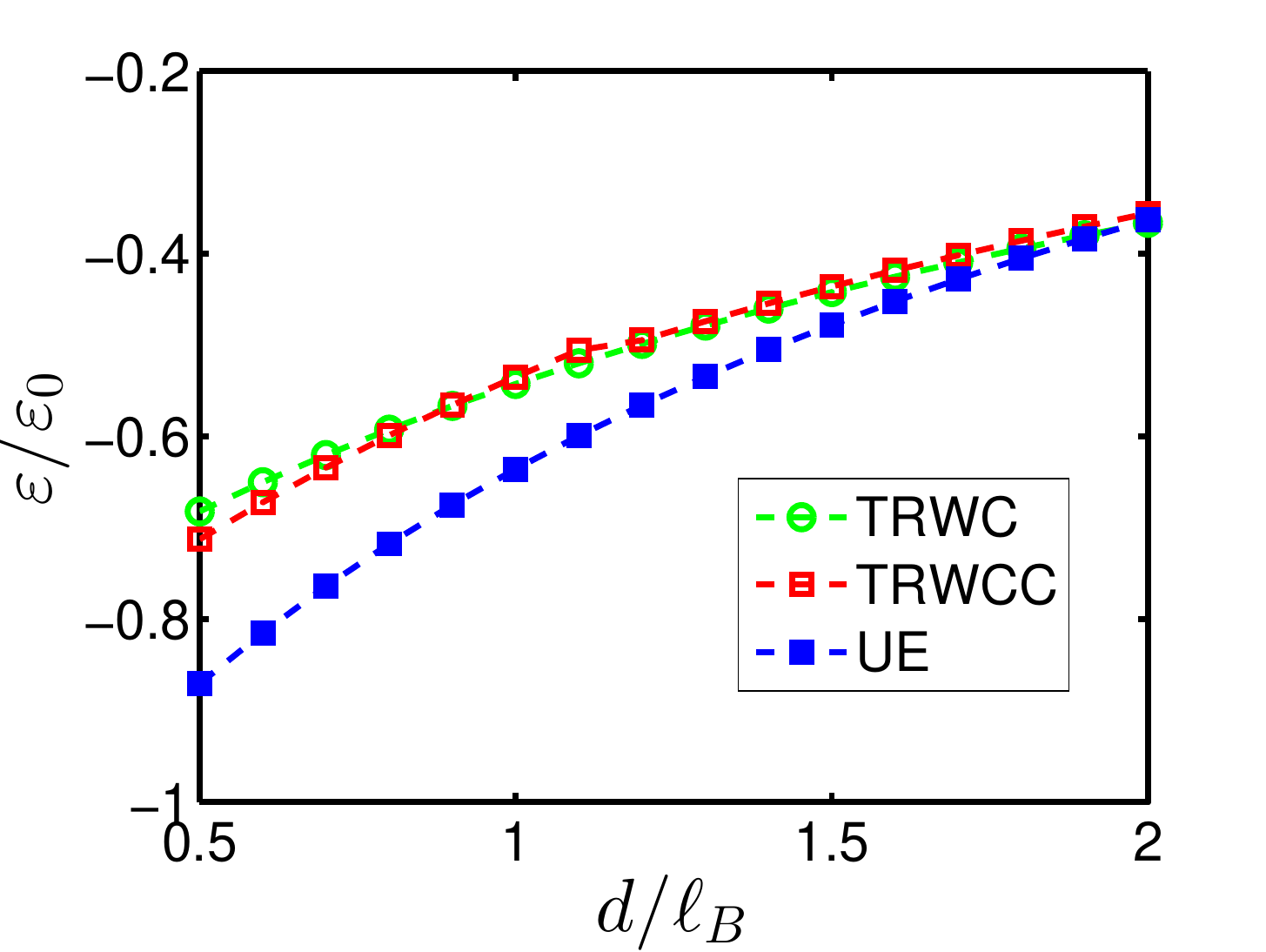}
        }%
\\ 
        \subfigure[]{%
            \label{Genc055-n2}
            \includegraphics[scale=0.4]{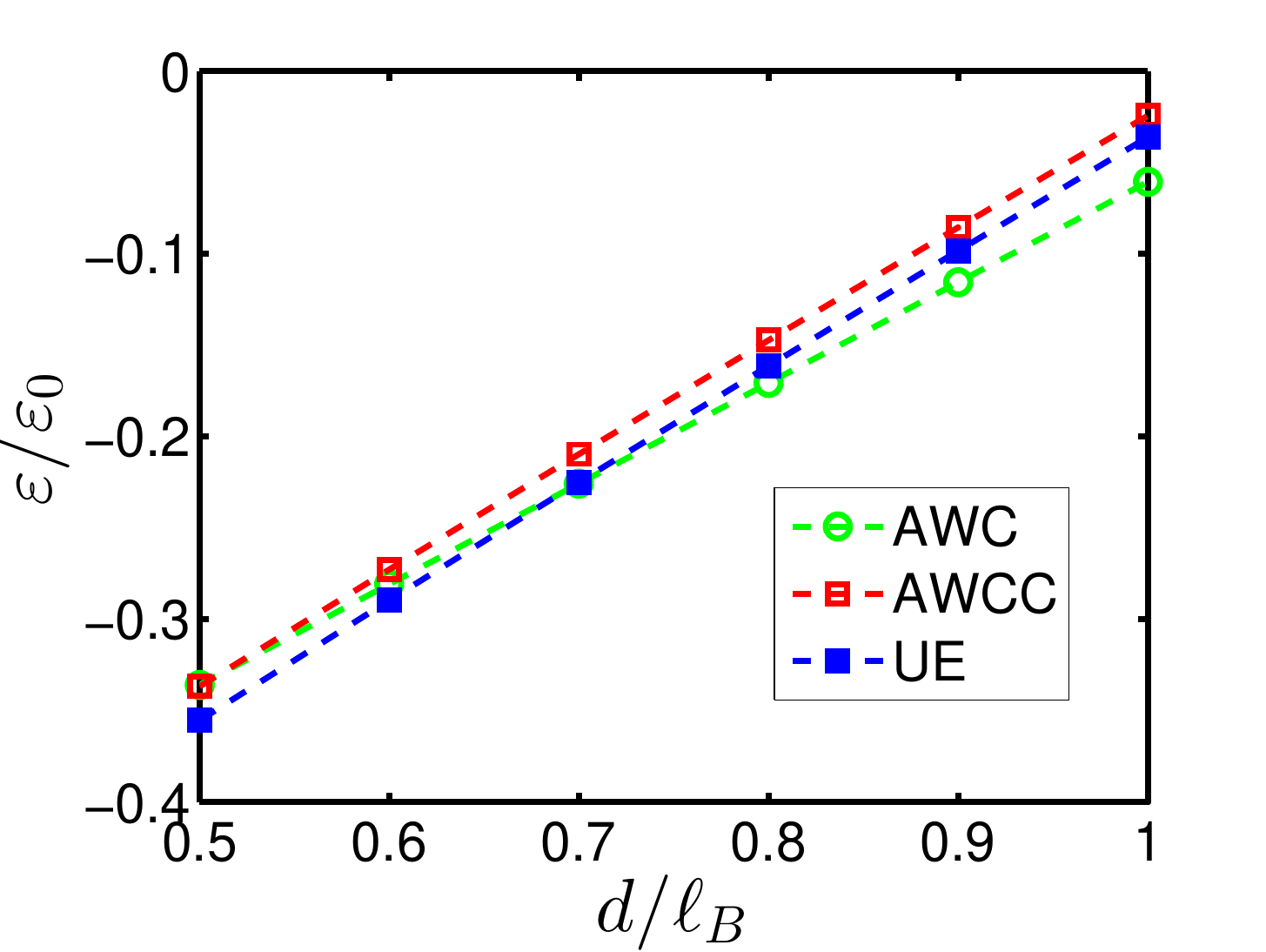}
        }%
    \end{center}
    \caption{%
        Comparison of energies per pair (in units of $\varepsilon_0=e^2/\epsilon\ell_B$) at two different Landau levels and filling factors as a function of interlayer separation. (a) The energy of triangular Wigner Crystal (TRWC) state vs. Triangular Coherent Wigner Crystal state (TRWCC) and
        uniform density excitonic state (UE) at Landau level $N=0$ and $\nu_T=0.05$. (b) The energy of Anisotropic Wigner Crystal State (AWC) compared with
        Coherent Anisotropic Wigner Crystal State (AWCC) and UE at $N=2$ and $\nu_T=0.55$. Here the anisotropy parameter is $\gamma=0.6$.
     }%
\label{comparison02}
\end{figure}

Assuming the Fourier transform of density matrices are only non-zero at points of reciprocal lattice space of a crystal of choice we can calculate the self-energy matrix ${\mathcal{M}}$ for a finite number of RLV's. This approximation is valid since the density matrix vanishes as we approach scales close to lattice constant in real space. Depending on the type of state we start with an initial guess for the density matrices and we find the converged solutions of the Eq. (\ref{eq-motion}). We have realized that most of our calculations converge with 16 RLV shells.

The lattice types that we choose are square, triangular and oblique lattices. In all these cases we choose the unit vectors in a way that there is only one carrier per unit cell. In the case of oblique lattice we choose the primitive lattice vectors ${\bf a}_1=\{a,b/2\}$ and ${\bf a}_2=\{0,b\}$ in which $a=\sqrt{2\pi/\nu\gamma}$ and the ratio $\gamma=b/a$ is a measure of \textit{anisotropy} of the lattice. Note that triangular lattice is a special case for $\gamma \approx 1.15$. The stripe states are in principle obtained by $\gamma \rightarrow 0$ or $\gamma\rightarrow \infty$.

We present our results for triangular and anisotropic crystal states for few values of $\gamma$. We point out that the energy of the square lattice has been found to be higher than any other lattice type in almost all the phase diagram.

For WC states as we mentioned before, although there is still considerable attraction between electrons and holes the interlayer coherence does not exist: $\rho_{eh}=0$. For such states we choose the density matrix so that electrons and holes form dipoles that are positioned on the chosen crystal sites: $\rho_{ee}(\bQ)=\rho_{hh}(\bQ) \neq 0$ in which $\bQ$ is an RLV. In this way the attraction between the electrons and holes will be maximum and the state would be lowest in energy.

For WCC states electrons and holes are paired and there is a quantum coherence between the two layers: $\rho_{eh}({\bf Q})\neq 0$. For such state we also choose to consider coherent states in which $\rho_{ee}=\rho_{hh}$ again to achieve lowest energy.

The structure of lattices then categorizes all states into incoherent triangular Wigner crystal (TRWC), coherent triangular Wigner crystal (TRWCC), incoherent anisotropic Wigner crystals (AWC) and finally coherent anisotropic Wigner crystal (AWCC).

Particle-hole symmetry maps double layer electron-hole system with total filling factor $\nu_T$ into another double layer with $2-\nu_T$. This implies the phase diagram to be symmetric around $\nu_T=1$ which is satisfied by our numerical results. Also based on Eqs. (\ref{eigval-eq}) and (\ref{green-eq}) the following sum rule holds:
\eqa
\sum_\bQ\left[|\rho_{ee}(\bQ)|^2+|\rho_{eh}(\bQ)|^2\right]=\rho_{ee}(0)=\nu.
\eqe
which is also satisfied for all our solutions up to order $10^{-7}$.

\textit{Phase Diagram}: We find the states that minimize the HF energy. In general for most of the separations $d\lesssim\ell_B$ we find that the UE state has the lowest energy compared to any crystalline state. In other parts of the phase diagrams we have not been able to find any type of coherent crystal that is lowest in energy throughout the whole phase diagram for $N=0,1,2,3$. We are demonstrating this in Fig. \ref{comparison02} which show a comparison of the energies of three UE, WC and WCC states with various crystal structures for sample partial filling factors and Landau levels (see below).


\begin{figure}[h]
\begin{center}
            \includegraphics[scale=.4]{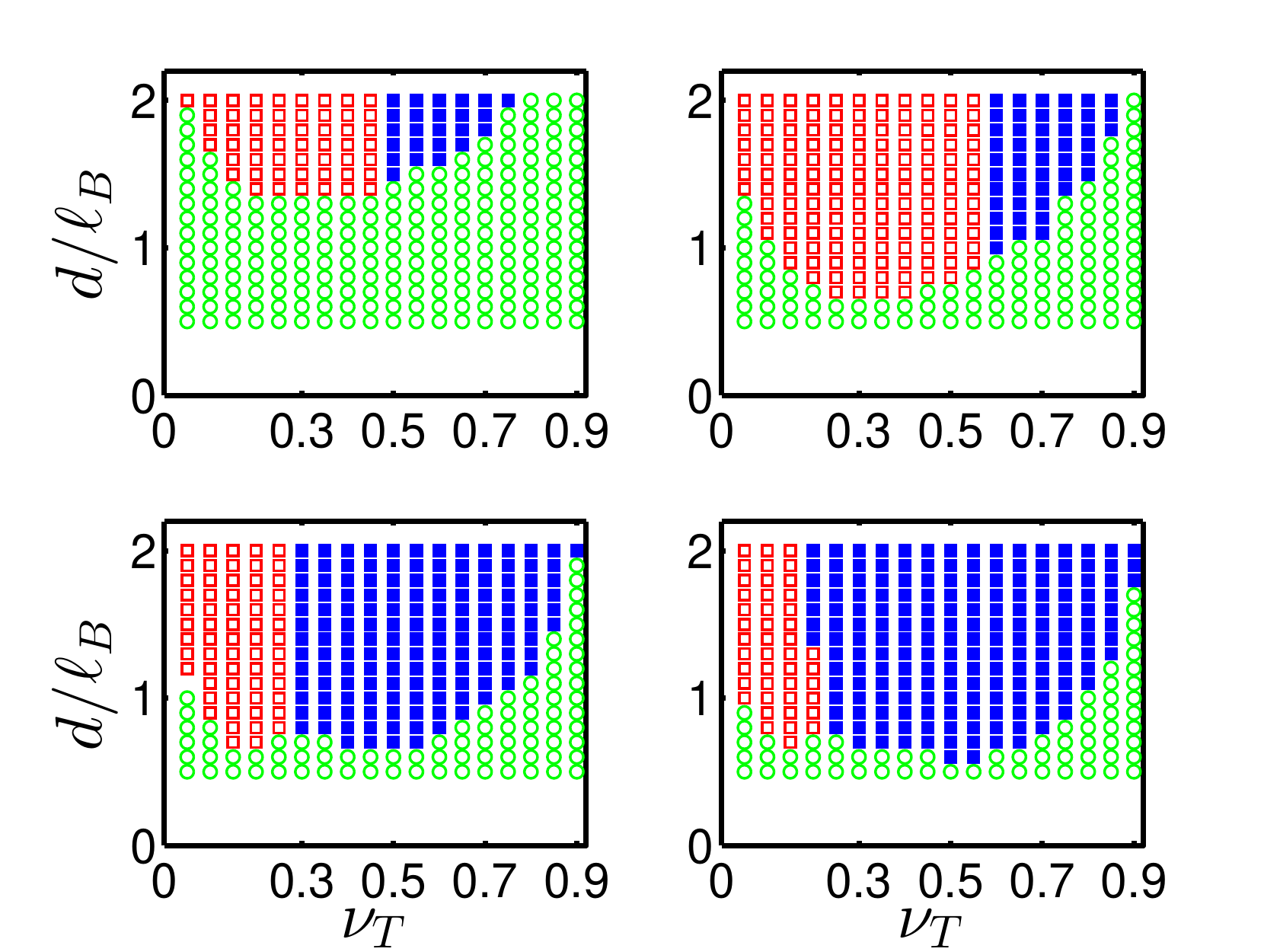}
\end{center}
\caption{Phase diagram for double sheet electron-hole graphene system at Landau levels $N=0,1,2,3$ as indicated in the plots.
        States are uniform excitonic condensate (open circles), triangular Wigner crystals of dipoles (open squares) and
        anisotropic wigner crystals (filled squares). The anisotropic states vary as a function of filling factor and Landau level, not distinguished in this pictures (See Fig.\ref{gamma}).}
        \label{Gphase}
\end{figure}

\begin{figure}[t]
\includegraphics[scale=0.4]{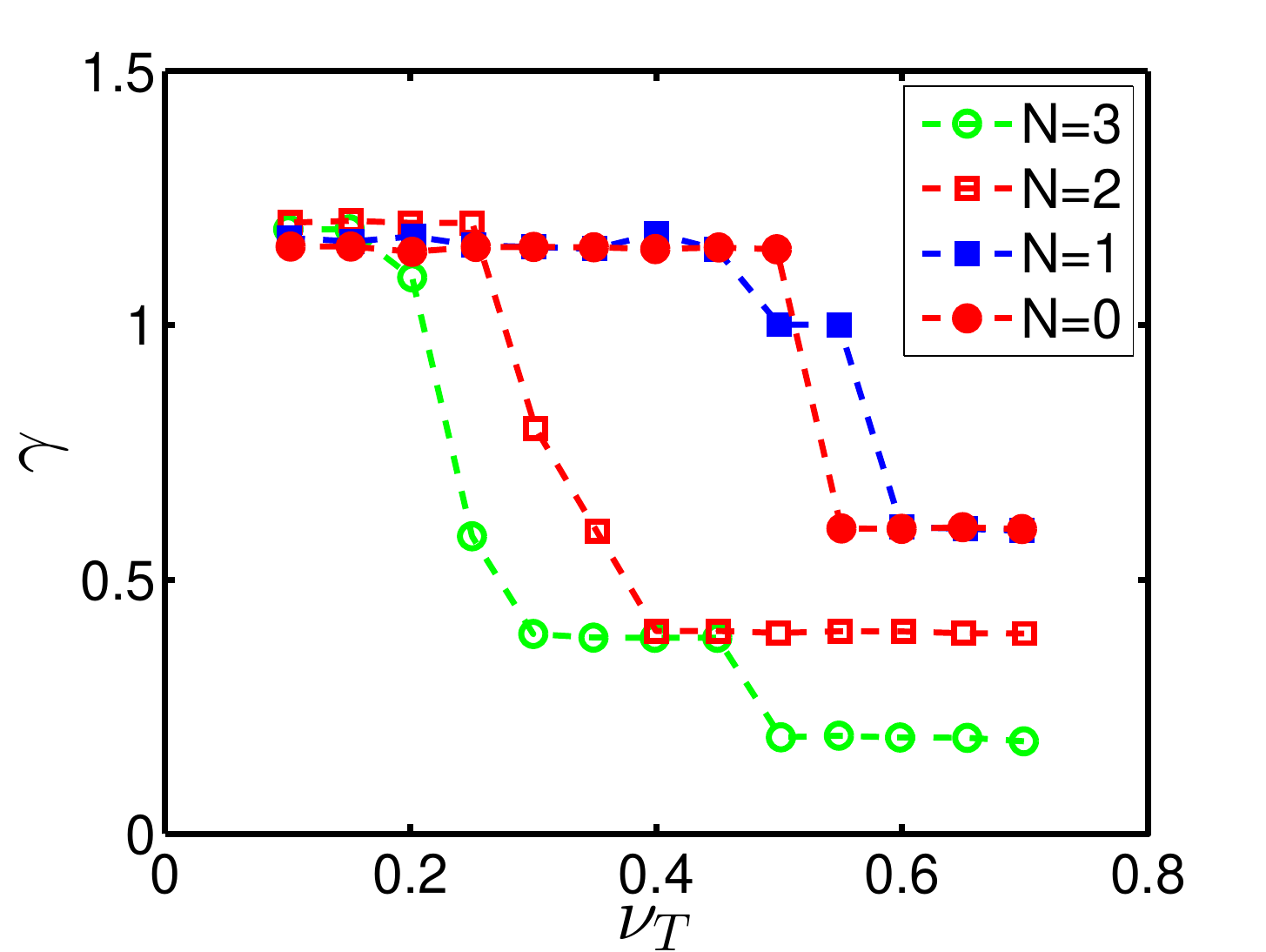}
\caption{Values of anisotropy parameter minimizing the energy of anisotropic wigner crystal state of dipoles (AWC) as a function of filling factor
for different Landau levels ($d=1.2\ell_B$).}
\label{gamma}
\end{figure}

At $N=0$ most of the WC states occur at  $d/\ell_B \gtrsim 1.2$. Fig. \ref{Gphase}, upper left shows the ground state phase diagram at lowest Landau level. Inside WC region and for $\nu_T \gtrsim 0.5 $ we see that there is a change from triangular into anisotropic (AWC) state. Inside the anisotropic state our method  is not capable of finding the exact value for $\gamma$ that minimizes the HF energy however sampling of a wide range of values $0.2 \leq \gamma \leq 2$ shows the ground state anisotropy is of the order of $\gamma \approx 0.6 $ for $0.5< \nu_T < 0.8$ at $d/\ell_B=1.2$. The value of $\gamma$ for interlayer separations close to this value does not change within our sample point accuracy. This behavior is the same in all the phase diagram. In Fig. \ref{gamma} we show the approximate value of $\gamma$ for ground states in different Landau levels at $d/\ell_B=1.2$.

\begin{figure}[t]
\includegraphics[scale=0.45]{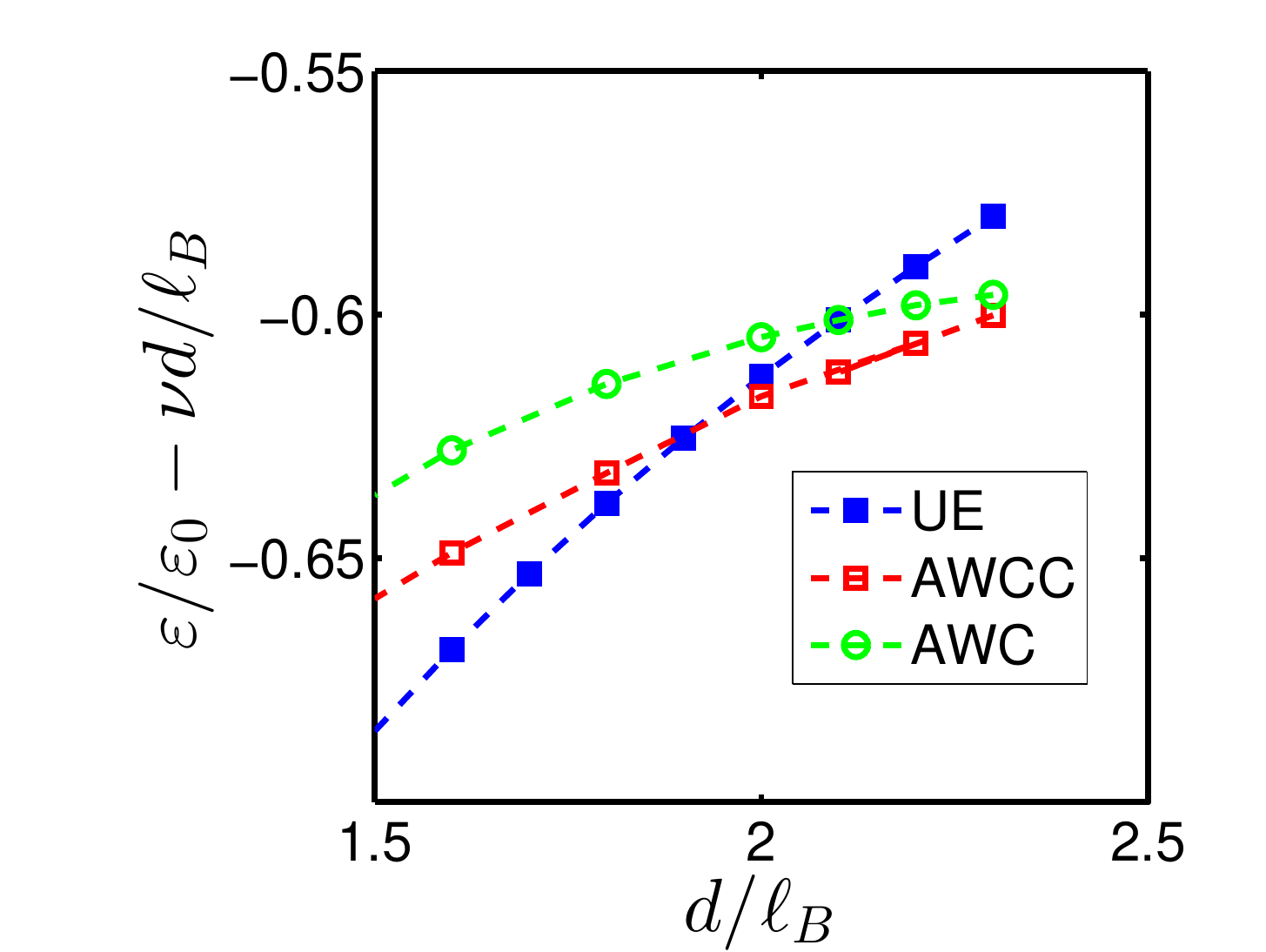}
\caption{Comparison of energy of the HF states (in units of $\varepsilon_0=e^2/\epsilon\ell_B$) for $\nu_T=2\times 0.23$ at lowest Landau level as a function interlayer separation. The states are uniform excitonic (UE), coherent crystalline with high anisotropy ($\gamma \approx 12$) indicated by AWCC and incoherent anisotropic (AWC) with the same value of anisotropy parameter.}
\label{stripe-energy}
\end{figure}

During the past investigations excitonic (coherent) states with broken translational symmetry states have been found to be the ground states\cite{Chen1991} in lowest Landau level in mean-field approximation. At first glance this seems to contradict our results however in those mean-field approximations only unidirectional (stripe) states were considered. In our calculations it is technically impossible to find those exactly unidirectional states however we have found out that for highly modulated stripe states indeed we obtain the excitonic states to lower the energy of the same state without coherence (see Fig. \ref{stripe-energy}). This indicates the fact that those findings in the past were only limited to a smaller selection of crystal structures. What we find out here is that those states are indeed higher in energy than certain incoherent crystalline structures.

For $N=1$ the anisotropy in the dipolar Wigner crystal ground state become more frequent in the phase space as can be seen in Fig. \ref{Gphase}, upper right. The phase boundary clearly has moved to smaller $d/\ell_B$ compared to the $N=0$ phase diagram. For filling factors $\nu_T \lesssim  0.6 $ and $d/\ell_B \gtrsim 0.8 $ the ground state is a triangular Wigner crystal. The crystal states become more anisotropic for $\nu_T \gtrsim 0.6 $. A careful comparison between energies of various WC and WCC crystal structures also shows no WCC state is a ground state in our HF approximation.

For $N=2,3$ the anisotropic WC state continues to advance into smaller filling factors and layer separations in phase diagram as can be seen in Fig. \ref{Gphase}, lower left and right. Also as illustrated by a sample filling factor in Fig. \ref{Genc055-n2} the WCC states do not show lower energy compared to WC states anywhere in our phase diagram. On the other hand for this high Landau levels WC ground states show more anisotropy in wider range of filling factors and layer separations.

\begin{figure}[t]
\includegraphics[scale=0.4]{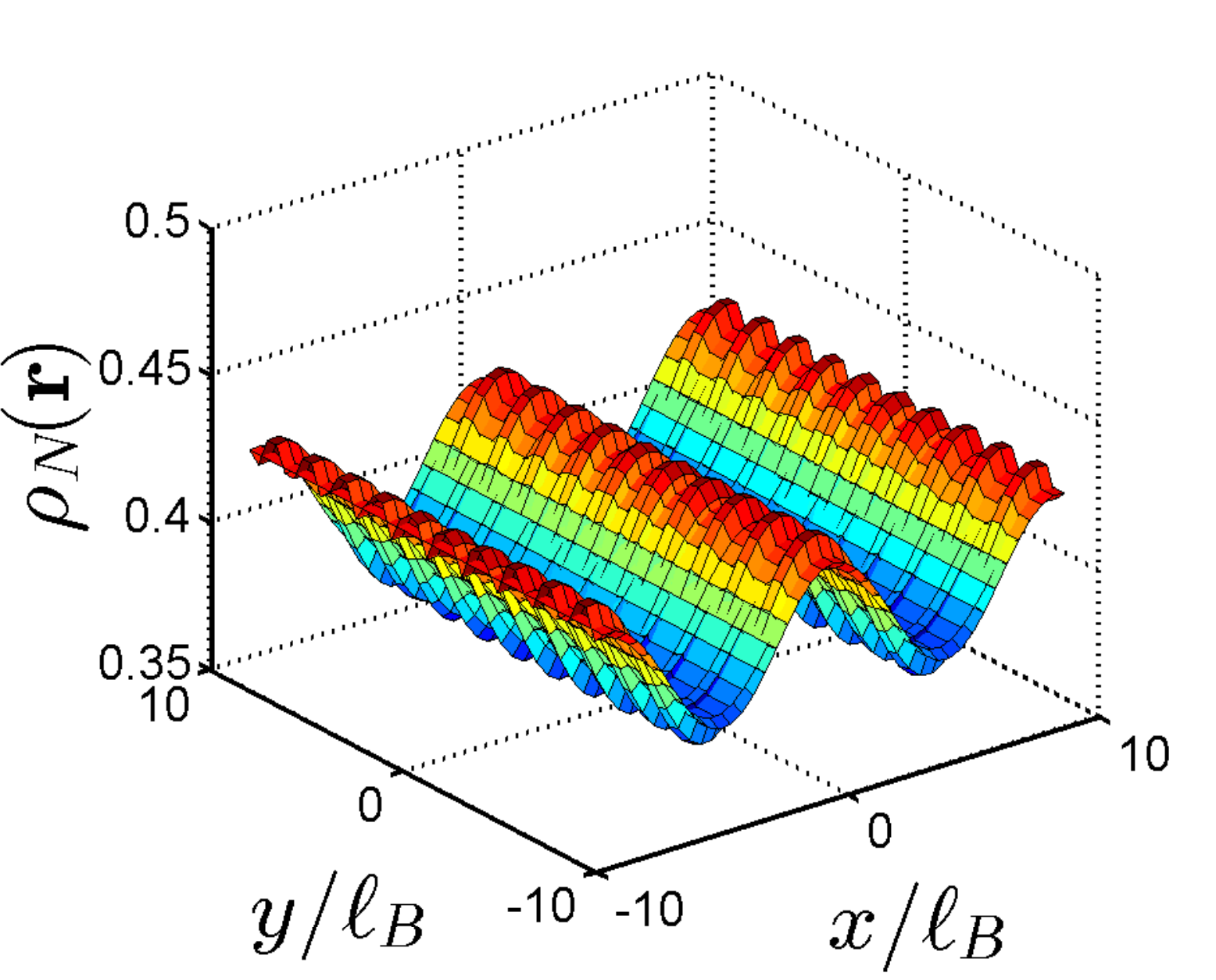}
\caption{Electron density (in units of $1/2\pi\ell_B^2$) profile for anisotropic incoherent Wigner crystal state (AWC) at $\nu_T=0.8$ and $d=1.2\ell_B$ in Landau level $N=3$. Here the anisotropy parameter is $\gamma=0.2$.}
\label{density-nu035-d102-n3}
\end{figure}

The AWC ground state anisotropy at higher Landau levels is to the degree that they almost resemble stripe states or more accurately \textit{modulated stripe} states. In these states stripes have periodic modulations. An example of such state is presented in Fig. \ref{density-nu035-d102-n3} for $\gamma=0.2$.

Throughout our investigations we found that states with both broken translational symmetry and U(1) symmetry are higher in energy than states with only one symmetry broken. This means in mean-field approximation those two symmetries break in crossing one single boundary, from one region to the other (by changing $d/\ell_B$). Note that this result means the translational symmetry breaks but U(1) symmetry is restored upon crossing the boundary in one direction (increasing $d/\ell_B$), that is why "breaking of both symmetries at the same time" is not an accurate description of the situation here. On the other hand quenching of the kinetic energy of charged particles into one Landau level is well known to affect their dynamics in a peculiar way\cite{Moon1995}. In our electron-hole system this translates to the fact that local exciton phase change and local density change are not completely independent. At small layer separation inside the UE phase, the excitons have established a uniform phase throughout the whole system which requires a uniform density development as well. This signals the fact that in low lying Landau levels a non-uniform density requires a non-uniform profile of the phase of the excitons. This modulation of the phase then will cost exchange energy compared to incoherent state where the phase is zero. We speculate the root of our numerical findings is connected to this fact although further investigation is necessary which is out of the scope of this work.


\begin{figure}[ht!]
            \includegraphics[scale=.4]{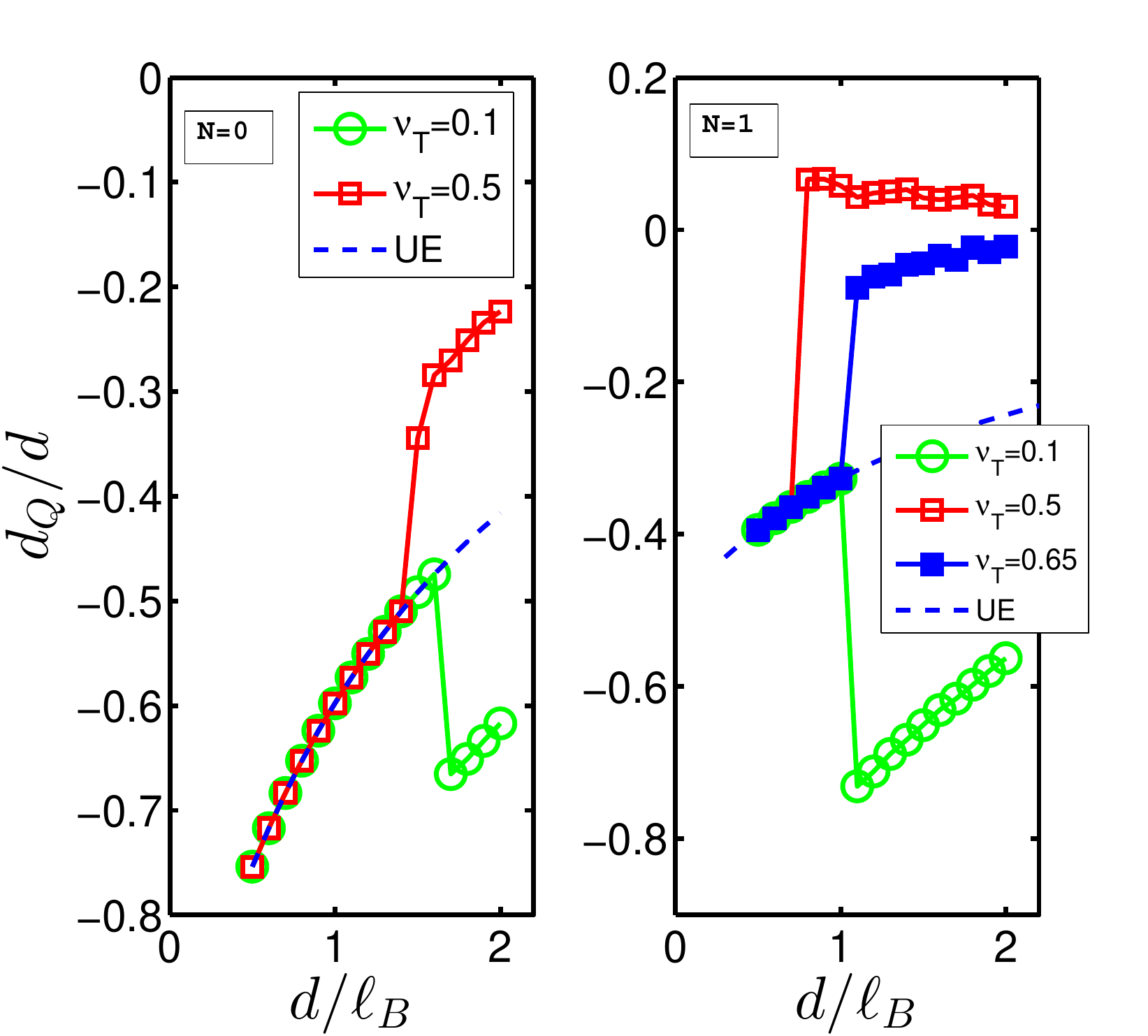}
    \caption{Ratio of Quantum Capacitance Length to interlayer separation $(d_Q/d)$ for different Landau levels $N,\nu_T$. This ratio can be calculated analytically for uniform excitonic state (UE, Dashed line). All the numerically calculated values for UE state agree with the analytic results. Solid lines are only for guide.}
    \label{GQCL-n01}
\end{figure}
\begin{figure}[ht!]
            \includegraphics[scale=.4]{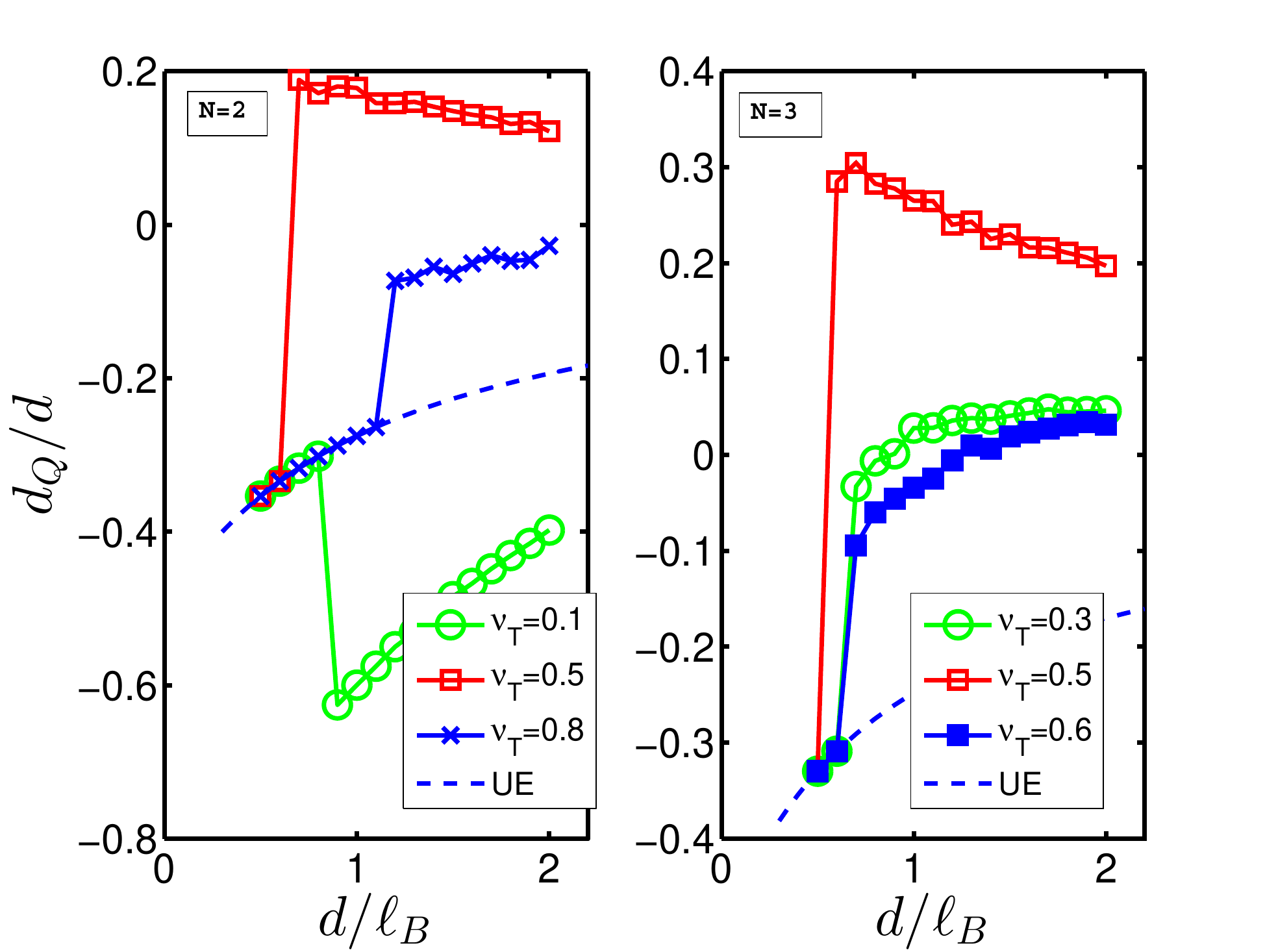}
    \caption{Ratio of Quantum Capacitance Length to interlayer separation $(d_Q/d)$ for different Landau levels $N,\nu_T$. This ratio can be calculated analytically for uniform excitonic state (UE, Dashed line). All the numerically calculated values for UE state agree with the analytic results. Solid lines are only for guide.}
     \label{GQCL-n23}
\end{figure}


\textit{Quantum Capacitance}: As ground state evolves across the phase diagram the quantum capacitance length calculated using Eq. (\ref{QCL-energy}) also shows a change in behavior. Figures \ref{GQCL-n01} and \ref{GQCL-n23} show our main numerical results for few sample filling factors. In all these results we see a jump in the value of $d_Q/d$ ratio as the layer separation increases due to the ground states changing from UE to WC state. The overall accuracy of our results are of the order of $10^{-6}$.

For states in lowest Landau level we see in almost all filling factors QCL is negative indicating the enhancement of capacitance. Note that for stability reasons $d^*\ge 0$ which means $d_Q/d \ge -1$ must hold. At dilute filling factor of $\nu_T=0.1$ in LLL (Fig. \ref{GQCL-n01}, left) and inside the crystalline phase $d^*$ is reduced to values close to -0.8! which means a reduction by about 70 to 60 percent depending on the layer separation. This indicates a giant capacitance compared to geometrical value.



The fate of the QCL at LLL at very small separations is determined by an analytic calculation based on equations (\ref{QCL-energy})-(\ref{B}) which gives a finite negative value $d_Q/d\rightarrow -1$ or infinite capacitance as $d\rightarrow 0$. This is because for two overlapping layers of opposite charges the dipole-dipole interaction vanishes. At separations close to zero as can be seen in all figures \ref{GQCL-n01} and \ref{GQCL-n23} our numerical results inside the UE phase are very close to analytic results. This is because in an exact solution of HF equations for such uniform state at finite layer separations the self-energy matrix elements are all zero except for $\bQ=0$.

Figure \ref{GQCL-n23} shows the QCL at higher Landau levels. In this figure specially for higher partial filling factors ($N=2,3$ and $\nu_T=0.5$) we see that QCL becomes positive which indicates the capacitance is reduced compared to classical values. In fact this behavior can be seen in all four panels in figures \ref{GQCL-n01} and \ref{GQCL-n23}. This indicates the \textit{recovering} of the capacitance, by increasing the magnetic field. The physical reason for such recovery in our model can be understood by noticing the change of the ground state configuration of the system. At higher Landau levels and higher partial filling factors the many body state is approaching a uniform density with modulations as we explained in our earlier discussion of the phase diagram. This \textit{tight} configuration resists additional charges and so its capacitance is lower.

Another important conclusion from our capacitance results is that the magnetic field gives us a control knob for the capacitance of the graphene system. Whether the experimental situations require higher or lower values of capacitance the applied field strength can be tuned accordingly. This concept has been in fact proven useful in recognizing various fractional quantum Hall states in bilayer electron-hole systems\cite{Yacoby2014}.

\begin{figure}[t]
            \includegraphics[scale=.3]{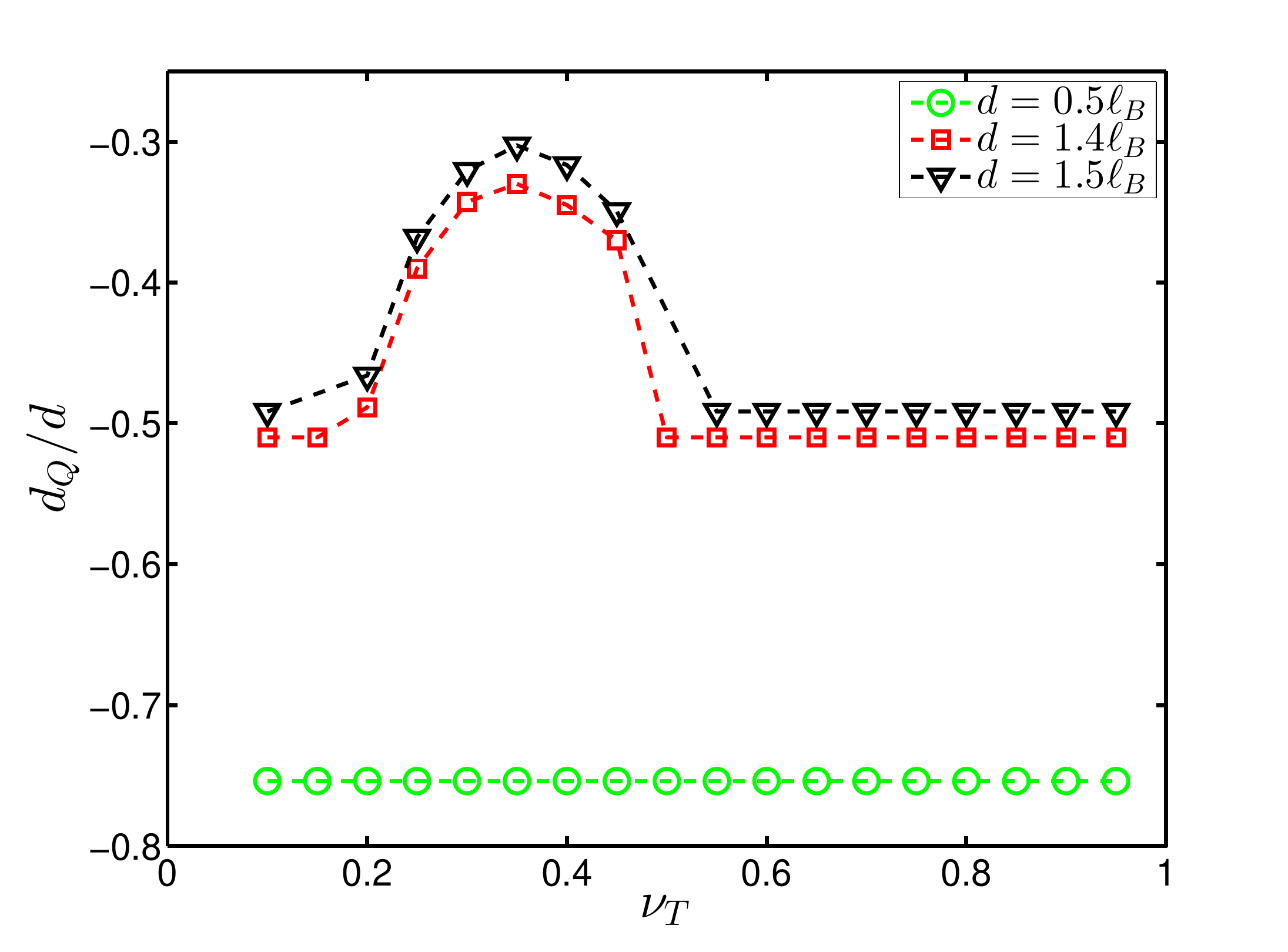}
    \caption{Ratio of Quantum Capacitance Length to interlayer separation $(d_Q/d)$ for different layer separations at $N=0$ as a function of electron filling factor. Dashed lines are only for guide.}
     \label{GQCL-nu}
\end{figure}

Finally note that the behavior of capacitance with respect to filling factor is non-monotonic. In all four Landau levels the QCL increases by increasing partial filling factor to $\nu_T\approx 0.3$ before it starts reducing again. This can be seen in figure \ref{GQCL-n01}, right panel. In this figure the QCL for $\nu_T=0.5$ at $N=1$ is higher than the curve for $\nu_T=0.65$. Same behavior is seen in both panels of Fig. \ref{GQCL-n23}. In Fig. \ref{GQCL-nu} we see the same results as a function of filling factor for $N=0$. Unfortunately our numerical procedure is not strong enough to be able to find stable solutions in all filling factors that is why we have been limited to few layer separations and filling factors for such curves. In this figure we can clearly see the peak of QCL at $\nu_T\sim 0.3$. At higher filling factors and by transition to UE state the QCL of course becomes flat. This flatness is because in mean field theory of uniform excitonic state the energy per dipole is only linearly dependent on the filling factor (see Eq. \ref{E-uniform}). That is why deep into the UE state at for example $d=0.5\ell_B$ we find a constant value for QCL as can be seen in Fig. \ref{GQCL-nu}.

The behavior of quantum capacitance length as a function of filling factor has been discussed before in Ref.\cite{Skinner2013}. In this work authors have derived the energy of the ground-state of the system at low filling factor using particle-hole symmetry and fitting to classical calculation of the energy of the Wigner crystal with a mean-field type first order quantum correction. The monotonic increase in QCL at low filling factors derived in this work is to some extent similar to the behavior seen in Fig.\ref{GQCL-nu}. However at higher filling factors or at very low filling factors where the crystal is expected to melt our results is clearly different in behavior.


\section{Conclusion}\label{section-last}
In this paper we have systematically studied the ground state of the balanced electron-hole double layer of graphene in strong magnetic field and the associated quantum corrections to its capacitance. We have focused on coherent and crystalline states in which U(1) and translational symmetries are broken respectively. By ignoring the inter-Landau level transitions we have approximately found the ground states for carrier occupations up to fourth lowest lying Landau levels in each layer.

In this article we showed that based on our calculations the anisotropic crystalline states take over a much larger part of the phase diagram at higher Landau levels (weaker magnetic fields). The anisotropic nature of these states is what one expects from the results of previous theoretical and experimental investigations\cite{Lilly-et-al} on the nature of quantum Hall liquids in higher Landau levels. According to those established theoretical results ground state phase of a single layer electron(hole) system is a stripe phase at higher Landau levels which may explain why two interacting stripe phases of electrons and holes form a stripe dipole state as is found in our numerical work.

On the other hand we showed that the application of strong magnetic field can tune the capacitance of the electron-hole system (or graphene-electrode system) to higher or lower than geometric values in agreement with experiment. We have shown that this effect is solely due to confinement of particles to lowest Landau levels and Coulomb interaction between electrons and holes.

We have also determined the phase diagram of the system and shown that the HF ground state energy of crystalline structures with exciton phase coherence is always higher than that without coherence. We emphasize again that there is no transfer of electrons between the two layers. However the coherence in the phase of excitons defined as the phase of off-diagonal element of density matrix is still well defined and the energy of states with different phase distribution or with average zero phase can be different.

Conceptually one expects a non-monotonic behavior in the phase of the system because at dilute regime, the crystal phase disappears in the phase diagram as it is melted by quantum fluctuations. On the other hand close to $\nu_T=1$ the system is equivalent to an almost balanced electron-electron bilayer system which is established as a uniform $\nu_T=1$ quantum Hall state even at the limit of zero interlayer tunneling\cite{Fertig89}.

We would like to note that the phase diagram and capacitance behavior of the electron-hole system obtained here are valid only in the thermodynamic limit and zero temperature. Finite size effects or presence of defects in the dipole crystals may alter the energy of states and may increase the possibility of some kind of coherent Wigner crystal of excitons (meaning with average non-zero excitonic phase). Also it is worth noting that HF approximation applied in this work has been proved in the past to capture the majority of the quasiparticle lattice energies in systems confined in the low Landau levels since this confinement suppresses the screening. The above points however must be investigated more carefully.

Finally it is important to note that states of electron-hole system such as the topological texture lattice states or more generally states with valley coherence may affect the behavior of capacitance at certain filling factor range. This is another question that is still open for investigation.

\section{Acknowledgement}
This work was fully supported by National Science Foundation of the United States, NSF-DMR 1054020. The author is grateful to Dr. Yogesh Joglekar for his supervision and useful discussions and Dr. Brian Skinner for reading the manuscript and providing useful comments. The author is also grateful to Csaba T\H{o}ke, Ali Naji and Reza Asgari for having useful discussions with the author. Finally author would like to thank Oklahoma Supercomputing Center (OSCER) for providing computational resources for this work.


\begin{thebibliography}{999}
\bibitem{Skinner2013}B. Skinner and B.I. Shklovskii, Phys. Rev. B {\bf 87}, 035409 (2013).
\bibitem{Skinner2013a}B. Skinner, G.L. Yu, A. V Kretinin, A.K. Geim, K.S. Novoselov, and B.I. Shklovskii, Phys. Rev. B {\bf 88}, 155417 (2013).
\bibitem{Yu2013}G.L. Yu, R. Jalil, B. Belle, A.S. Mayorov, P. Blake, F. Schedin, S. V Morozov, L. a Ponomarenko, F. Chiappini, S. Wiedmann, \textit{et. al.}, Proc. Natl. Acad. Sci. U. S. A. {\bf 110}, 3282 (2013).
\bibitem{skinner2010}Brian Skinner and B. I. Shklovskii, Phys. Rev. B {\bf 82}, 155111 (2010).
\bibitem{E-H}Yogesh N. Joglekar, Alexander V. Balatsky, and S. Das Sarma, Phys. Rev. B {\bf 74}, 233302 (2006),
Oleg L. Bermana, Roman Ya. Kezerashvilia and Klaus Ziegler, Physica E: Low-dimensional Systems and Nanostructures {\bf 71}, pp. 7-13 (2015), S. De Palo, F. Rapisarda and Gaetano Senatore, Phys. Rev. Lett. {\bf 88}, 206401 (2002), Oleg L. Berman, Yurii E. Lozovik and Godfrey Gumbs, Phys. Rev. B {\bf 77}, 155433 (2008).
\bibitem{Butov2004} L. V. Butov, J. Phys.:Condens. Matter {\bf 16} (2004) R1577-R1613.
\bibitem{Fertig89}H. A. Fertig, Phys. Rev. B {\bf 40}, 1087 (1989).
\bibitem{stablished-state}I. B. Spielman, J. P. Eisenstein, L. N. Pfeiffer, and K. W. West
Phys. Rev. Lett. {\bf 84}, 5808 (2000), M. Abolfath, L. Radzihovsky, and A. H. MacDonald
Phys. Rev. B {\bf 65}, 233306 (2002), D. N. Sheng, Leon Balents, and Ziqiang Wang
Phys. Rev. Lett. {\bf 91}, 116802 (2003), John Schliemann, S. M. Girvin, and A. H. MacDonald
Phys. Rev. Lett. {\bf 86}, 1849 (2001), John Schliemann and A. H. MacDonald
Phys. Rev. Lett. {\bf 84}, 4437 (2000).
\bibitem{imbalance}Kun Yang, Phys. Rev. Lett. {\bf 87}, 056802 (2001), Yogesh N. Joglekar and Allan H. MacDonald
Phys. Rev. B {\bf 65}, 235319 (2002).
\bibitem{Chen1991}X. M. Chen and J. J. Quinn, Phys. Rev. Lett. {\bf 67},  895 (1991).
\bibitem{Zheng-Fertig}Lian Zheng and H. A. Fertig, Phys. Rev. B {\bf 52}, 12282 (1995).
\bibitem{Cote1992}R. Cote, L. Brey and A.H. MacDonald, Phys. Rev. B {\bf 46}, 10239 (1992).
\bibitem{Narasimhan1995} Subha Narasimhan and Tin-Lun Ho, Phys. Rev. B {\bf 52}, 12291 (1995).
\bibitem{Csaba2014} Csaba T\H{o}ke and Vladimir I. Fal'ko, Phys. Rev. B {\bf 90}, 035404 (2014).
\bibitem{Wigner}E. Wigner, Phys. Rev. {\bf 46}, 1002 (1934).
\bibitem{Geim2013}A. K. Geim and I. V Grigorieva, Nature {\bf 499}, 419 (2013).
\bibitem{Yoshioka1990}D. Yoshioka and A. H. Macdonald, J. Phys. Soc. Jpn. {\bf 59}, 4211 (1990).
\bibitem{Zhang2007}C.-H. Zhang and Y.N. Joglekar, Phys. Rev. B {\bf 75}, 245414 (2007).
\bibitem{CastroNeto2009}A. H. Castro Neto, N.M.R. Peres, K.S. Novoselov, and a. K. Geim, Rev. Mod. Phys. {\bf 81}, 109 (2009).
\bibitem{method}For basic formulation see for example:  R. Cote, L. Brey and A.H. MacDonald, Phys. Rev. B {\bf 46}, 10239  (1992) also see Ref.\cite{Zhang2007}.
\bibitem{Dahal2006}H.P. Dahal, Y.N. Joglekar, K.S. Bedell, and A.V. Balatsky, Phys. Rev. B {\bf 74}, 233405 (2006).
\bibitem{Eisenstein1994}J. P. Eisenstein, L. N. Pfeiffer, and K. W. West, Phys. Rev. B {\bf 50}, 1760 (1994).
\bibitem{Moon1995}K. Moon, H. Mori, K. Yang, S.M. Girvin, A. H. MacDonald, L. Zheng, D. Yoshioka and S.C. Zhang, Phys. Rev. B {\bf 51}, 5138 (1995).
\bibitem{Yacoby2014}A. Kou, B. E. Feldman, A. J. Levin, B. I. Halperin, K. Watanabe, T. Taniguchi, and A. Yacoby
Science {\bf 345}, 1250270 (2014).
\bibitem{Lilly-et-al} M.P. Lilly, K.B. Cooper, J.P. Eisenstein, L.N. Pfeiffer, and K.W. West, Phys. Rev. Lett. {\bf 82}, 394 (1999); R.R. Du, D.C. Tsui, H.L. Stormer, L.N. Pfeiffer, K.W. Baldwin, and K.W. West, Solid State Commun. {\bf 109}, 389 (1999); L. Brey and H. A. Fertig, Phys. Rev. B {\bf 62}, 10268 (2000).




\end{thebibliography}
\end{document}